\documentclass[journal]{IEEEtran}

\usepackage[utf8]{inputenc}
\usepackage{amsmath}
\usepackage{graphicx}
\usepackage{fullpage,marginnote,caption}
\usepackage{amsfonts}
\usepackage{amsmath}
\usepackage{bm}
\usepackage{fullpage}
\usepackage{amssymb}
\usepackage{graphicx}
\usepackage{fancyhdr,color}
\usepackage[utf8]{inputenc}
\usepackage[english]{babel}
\usepackage{subcaption} 
\usepackage[algoruled,boxed,lined]{algorithm2e}
\usepackage{tabularx}
\usepackage{siunitx}
\usepackage{booktabs}
\usepackage{adjustbox}
\usepackage{flushend}
\allowdisplaybreaks

\usepackage{nopageno} 
\usepackage{bbm} 
\usepackage{dsfont} 
\usepackage[normalem]{ulem} 

\usepackage{color,soul} 
\soulregister\cite7
\soulregister\ref7
\soulregister\pageref7

\usepackage[utf8]{inputenc}
\usepackage[english]{babel}

\newcommand{\real}{{\mathbb{R}}}

\graphicspath{Figures/ }
 
\usepackage{geometry}
\ifCLASSINFOpdf
\else
\fi

\graphicspath{ {Figures/} }

\begin{document}


%
\title{\vspace{5.4mm} Reinforcement Learning-based Fast Charging \\  Control Strategy for  Li-ion Batteries}
%
%
%

\author{Saehong Park, Andrea Pozzi, Michael Whitmeyer, Hector Perez, \\ Won Tae Joe, Davide M Raimondo, Scott Moura
\thanks{ Saehong Park, Michael Whitmeyer, Hector Perez, and Scott Moura are with the Energy, Controls and Applications Lab (eCAL) at the University of California, Berkeley, CA 94720, USA (E-mail: \{sspark,mwhitmeyer, heperez, smoura\}@berkeley.edu)}
\thanks{Andrea Pozzi and Davide M Raimondo is with University of Pavia, Corso Str. Nuova, 65, 27100, Pavia, Italy (E-mail: andrea.pozzi03@universitadipavia.it, davide.raaimondo@unipv.it)}
\thanks{Won Tae Joe is with LG Chem,  BMS Advanced SW Project Team, Yuseong-gu, Daejeon, 305-738, South Korea (E-mail: wontaejoe@lgchem.com)}}

\maketitle
\thispagestyle{plain}

\begin{abstract}

One of the most crucial challenges faced by the Li-ion battery community concerns the search for the minimum time charging without damaging the cells. This can fall into solving large-scale nonlinear optimal control problems according to a battery model. Within this context, several model-based techniques have been proposed in the literature. However, the effectiveness of such strategies is significantly limited by model complexity and uncertainty. Additionally, it is difficult to track parameters related to aging and re-tune the model-based control policy. With the aim of overcoming these limitations, in this paper we propose a fast-charging strategy subject to safety constraints which relies on a model-free reinforcement learning framework. In particular, we focus on the policy gradient-based actor-critic algorithm, i.e., deep deterministic policy gradient (DDPG), in order to deal with continuous sets of actions and sets. The validity of the proposal is assessed in simulation when a reduced electrochemical model is considered as the real plant. Finally, the online adaptability of the proposed strategy in response to variations of the environment parameters is highlighted with consideration of state reduction.

\end{abstract}


%
\IEEEpeerreviewmaketitle

\section{Introduction}
\label{sec:Intro}



Lithium-ion batteries are crucial technologies for electrified transportation, clean power systems, and consumer electronics. Although among all the different chemistries, Li-ion batteries exhibit promising features in terms of energy and power density, they still present limited capacity and long charging time. While the former is mostly related to the battery chemistry and design phase, the latter depends on the employed charging strategy. Within this context, the trade-off between fast charging and aging has to be taken into account. In fact, charging time reductions can be easily achieved by using aggressive current profiles which in turn may lead to severe battery degradation effects, such as Solid Electrolyte Interphase (SEI) growth and Lithium plating deposition. For this reason, several model-based optimal control techniques have been proposed in the literature with the aim of providing fast-charging while guaranteeing safety constraints.

The authors in \cite{klein2011optimal} formulate a minimum-time charging problem and use nonlinear model predictive control. Similarly, authors in \cite{torchio2015real} propose quadratic dynamic matrix control formulation to design an optimal charging strategy for real-time model predictive control. In the context of aging mechanism, the authors of \cite{perez2017optimal} have studied the trade-off between charging speed and degradation, based on an electro-thermal-aging model. The authors in \cite{pozzi2018film} consider minimizing film layer growth of the electrochemical model. Authors in \cite{suthar2014optimal,suthar2014optimalJES} derive an optimal current profile using a single particle model with intercalation-induced stress generation. The key novelty here is incorporating mechanical fracture, which can be a dominant mechanism in degradation and capacity fade. To ensure safety, a proportional–integral–derivative controller is proposed. On the other hand, the authors in \cite{zou2018electrochemical} synthesize a state estimation and model predictive control scheme for a reduced electrochemical-thermal model, in order to design health-aware fast charging strategy. The problem is formulated as a linear time-varying model predictive control scheme, with a moving horizon state estimation framework. In \cite{liu2017extended}, the authors exploit differential flatness properties of the single particle model to yield a computationally efficient optimal control problem, solved via pseudospectral methods. 

However, the exploitation of model-based charging procedure has to face some crucial challenges. (i) Every model is inherently subject to modeling mismatches and uncertainties. (ii) The most commonly used detailed models for Li-ion batteries are the electrochemical ones which typically contain hundreds or thousands of states, leading to a large-scale optimization problem. (iii) The model parameters drift as the battery ages. It is important to notice that most of the model-based strategies proposed in the literature rely on simplified electrochemical models (the few ones which implement full order models represent the boundary of what can be done in this area) and  almost none of them consider adaptability of the control strategy to variations in the parameters. In addition, electrochemical models present observability and identifiability issues \cite{moura2015estimation}, which often lead to the necessity of optimally designing the experiments which have to be conducted in order to properly estimate the parameters with a sufficiently high accuracy  \cite{pozzi2018optimal,park2018optimal}. All these issues can be addressed by using a charging procedure based on a model-free Reinforcement Learning (RL) framework \cite{sutton2018reinforcement}. An RL framework consists of an agent (the battery management system) which interacts with the environment (the battery) by taking specific actions (the applied current) according to the environment configuration (a.k.a. the state). The model-free property implies that the agent learns online the feedback control policy, directly from interactions with the environment, such as reward and state observation. Such policy is iteratively updated in order to maximize the expected long-term reward. Notice that, the reward has to be properly designed in order to make the agent learn how to accomplish the required task. 

Most RL algorithms can be classified in two different groups: tabular methods, e.g., Q-learning, SARSA, and approximate solutions methods which is also called  ``Approximate Dynamic Programming (ADP)''. While the former performs well only in presence of small and discrete set of actions and states, the latter can be used even with continuous state and action spaces solving the so-called ``curse of dimensionality''. On the other hand, the convergence of the former is proven under mild assumptions. However, no proof of convergence exists for the approximate methods in the general case. 
The recent success in several applications of RL based on deep neural networks as function approximators has greatly increased expectations in the scientific community \cite{mnih2015human,schulman2015trust, van2016deep,belletti2017expert}. From a control systems perspective, the design of RL algorithms involves feedback control laws for dynamical systems via optimal adaptive control methods \cite{lewis2012reinforcement}.
It is also important to notice that several works have been focused in developing safe-RL strategies, which are able to learn optimal control policy while guaranteeing safety constraints \cite{garcia2015comprehensive}, which are fundamental in the context of battery fast-charging.

In this paper, a fast-charging strategy subject to safety constraints, using a model-free reinforcement learning framework, is proposed for the first time to the knowledge of the authors in the context of Li-ion batteries. The use of such a methodology  enables adaptation to uncertain and drifting parameters. Moreover, the exploitation of ADP-based approaches allows one to mitigate the curse of dimensionality for large-scale nonlinear optimal control problems by adopting parameterized actor/critic networks.
In particular, we focus on  the Deep Deterministic Policy Gradient (DDPG) \cite{lillicrap2015continuous} algorithm, which is an actor-critic formulation suitable for the case of continuous actions space and includes deep neural networks as function approximators. The safety constraints are considered by including a penalty in the reward function in case of violation. The control technique is tested by considering a Single Particle Model with Electrolyte and Thermal (SPMeT) \cite{perez2017optimalJES} dynamics as the battery simulator. Two different scenarios are presented: in the first one all the states are assumed measurable from the agent, while in the second this assumption is dropped and only state of charge and temperature are considered available. The results show that the RL-agent is able to achieve high performance in both the scenarios. Finally, we examine the online adaptability of the proposed methodology in the case of varying parameters, i.e. degradation.

The paper is organized as follows. Section~\ref{sec:RL_framework} briefly presents the reinforcement learning approach. Section~\ref{sec:BattModel} describes the battery models and control problem formulation. Section~\ref{sec:Sim_results} presents case study with simulation results. In Section~\ref{sec:Conclusion}, we summarize our work and provide perspectives on future work.

\vspace{-1ex}
\section{Reinforcement Learning Approach}
\label{sec:RL_framework}


\begin{figure}[t]
\centering
\includegraphics[ width=0.35\textwidth]{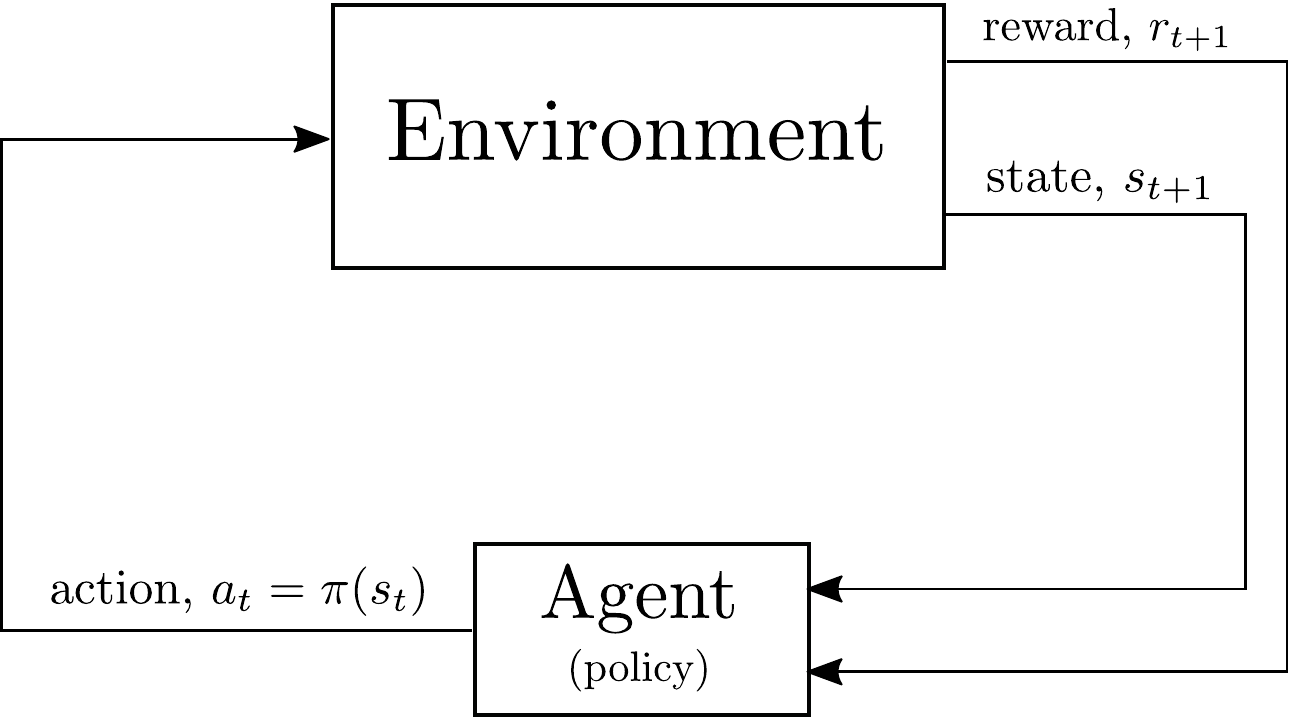}
\caption{Reinforcement Learning framework.}
\vspace{-4ex}
\label{fig:RL_framework}
\end{figure}

In this section a standard reinforcement learning setup is presented, along with the main feature of Approximate Dynamic Programming and actor-critic algorithm. \vspace{-2ex}

\subsection{Markov Decision Process, Policy and Value Functions}

In the reinforcement learning framework shown in Fig.~\ref{fig:RL_framework}, we seek the best policy that will maximize the total rewards received from the environment $E$ (i.e. plant). At each time step $t \in \mathbb{R}^+$ the environment exhibits state vector, $\bm{s}_t\in \mathcal{S}$, where $\mathcal{S}$ is the state space, the control policy (a.k.a. agent) observes the states $\bm{s}_{t}$ and picks an action $\bm{a}_{t}\in \mathcal{A}$, with $\mathcal{A}$ being the action space. This action is executed on the environment, whose state evolves to $\bm{s}_{t+1}\in \mathcal{S}$, according to the state-transition  probability  $p(\bm{s}_{t+1}|\bm{s}_{t},\,\bm{a}_{t})$, and the agent receives a scalar reward $\bm{r}_{t+1}=r(\bm{s}_t,\bm{a}_t)$. The policy is represented by $\pi$ which maps the state to the action and can be either deterministic or stochastic. The total discounted reward from time $t$ onward can be expressed as:
\begin{align}
	R_{t} 
	       & =\sum_{k=0}^{\infty} \gamma^{k}r(\bm{s}_{t+k},\bm{a}_{t+k})
\end{align}
where $\gamma \in [0,1]$ is the discounting factor.


The \textit{state value function}, $V^{\pi}(\bm{s}_t)$ is the expected total discounted reward starting from state $\bm{s}_t$. In the controls community, this is sometimes called the cost-to-go, or reward-to-go. Importantly, note the value function depends on the control policy. If the agent uses a given policy $\pi$ to select actions starting from the state $\bm{s}_t$, the corresponding value function is given by:
\begin{align}
	V^{\pi}(\bm{s}_t) \doteq \mathbb{E}_{r_{i>t},\,s_{i>t}\sim E,\,a_{i\geq t}\sim \pi} \Big[ R_{t} \ | \  \bm{s}_{t} \Big]
    \label{eqn:Vfct}
\end{align}



Then, the optimal policy $\pi^{*}$ is the policy that corresponds to the maximum  value of the value function
\begin{align}\label{eqn:Vmax}
	\pi^{*} = \text{arg}\max_{\pi} V^{\pi}(\bm{s}_t)
\end{align}
The solution of \eqref{eqn:Vmax} is pursued by those methods  which follow the Dynamic Programming (DP) paradigm. Such paradigm assumes a perfect knowledge of the environment E (i.e, the state-transition probability as well as the reward function are known).

The next definition, known as the ``Q-function,'' plays a crucial role in model-free reinforcement learning.  Consider the \textit{state-action value function}, $Q^{\pi}(\bm{s}_t,\bm{a}_t)$, which is a function of the state-action pair and returns a real value. This Q-value corresponds to the long-term expected return when action $\bm{a}_t$ is taken in state $\bm{s}_t$, and then the policy $\pi$ is followed henceforth. Mathematically,
\begin{align}
    Q^{\pi}(\bm{s}_t,\bm{a}_t)  \doteq \mathbb{E}_{r_{i>t},\,s_{i>t}\sim E,\,a_{i> t}\sim \pi}\Big[ R_{t} \ | \ \bm{s}_{t},\, \bm{a}_{t} \Big] \label{eqn:Qvalue}
\end{align}
The state-action value function can be expressed as Bellman equation, such as:
\begin{align}
\begin{split}
    Q^{\pi}(\bm{s}_t,\bm{a}_t)  =& \mathbb{E}_{r_{i>t},\,s_{i>t}\sim E} \Big[r(\bm{s}_t,\bm{a}_t) +\\& \gamma\mathbb{E}_{a_{t+1}\sim \pi}\big[  Q^{\pi}(\bm{s}_{t+1},\bm{a}_{t+1})\big]\Big] 
    \end{split}
\end{align}

The optimal Q-function $Q^{{*}}(\bm{s}_t,\bm{a}_t)$ gives the expected total reward received by an agent that starts in $\bm{s}$, picks (possibly non-optimal) action $\bm{a}_t$, and then behaves optimally afterwards. $Q^{{*}}(\bm{s}_t,\bm{a}_t)$ indicates how good it is for an agent to pick action $\bm{a}_t$ while being in state $\bm{s}_t$. Since $V^{{*}}(\bm{s}_t)$ is the maximum expected total reward starting from state $\bm{s}_t$, it will also be the maximum of $Q^{{*}}(\bm{s}_t,\bm{a}_t)$ over all possible actions $\bm{a}_t\in \mathcal{A}$
\begin{align}
	V^{{*}}(\bm{s}_t) = \max_{\bm{a}_t\in \mathcal{A}} Q^{{*}}(\bm{s}_t,\bm{a}_t)
\end{align}
If  the optimal Q-function is known, then the optimal action $\bm{a}^*_t$ can be extracted by choosing the action $\bm{a}_t$ that maximizes $Q^{{*}}(\bm{s}_t,\bm{a}_t)$ for state $\bm{s}_{t}$ (i.e. the optimal policy $\pi^*$ is retrieved),
\begin{equation}\label{eqn:Qmax}
	\bm{a}^*_t = \text{arg}\max_{\bm{a}_t \in \mathcal{A}} Q^{*}(\bm{s}_t,\bm{a}_t)
\end{equation}
without requiring the knowledge of the environment dynamics.
\vspace{-2ex}

\subsection{Tabular Methods and Approximated Solutions}
In a model-free framework, the Q-function can be learned directly  from the interaction with the environment, by means of the reward collected over time. Within this context two different approaches can be considered: tabular methods and ADP \cite{sutton2018reinforcement,bertsekas2005dynamic}. The former store the Q-function as a table whose entrance are the states and the actions, while the latter uses parameterized Q-function using Value Function Approximation (VFA). The main advantage of ADP relies in its ability of solving the so-called curse of dimensionality, which is a negative feature of both DP and reinforcement learning strategies based on tabular methods \cite{powell2007approximate}. In particular, the curse of dimensionality consists on the exponential rise in the time and space required to compute a solution to an MDP problem as the dimension (i.e. the number of
state and control variables) increase \cite{rust1997using}. Due to such issue the use of both DP and tabular methods is limited to the context of small and discrete action and state spaces.

Let consider the following approximation
\begin{align}
    Q^\pi(\bm{s}_t,\bm{a}_t)\approx Q(\bm{s}_t,\bm{a}_t|\theta^{Q^\pi})
\end{align}
The idea used by ADP methods to solve the curse of dimensionality is to seek for the optimal parameters vector $\theta^{Q^\star}$ instead directly for the Q-function $Q^\star(\bm{s}_t,\bm{a}_t)$, thus  reducing significantly the size of the optimization problem. Several function approximators can be employed, e.g. linear approximators, neural networks, kernel-based functions. 
One of the most famous example of ADP using deep neural networks as VFAs is given in \cite{mnih2015human}, where the deep Q-learning algorithm is proposed.

\begin{figure}[t]
\centering
\includegraphics[trim = 50mm 15mm 50mm 15mm, clip, width=0.4\textwidth]{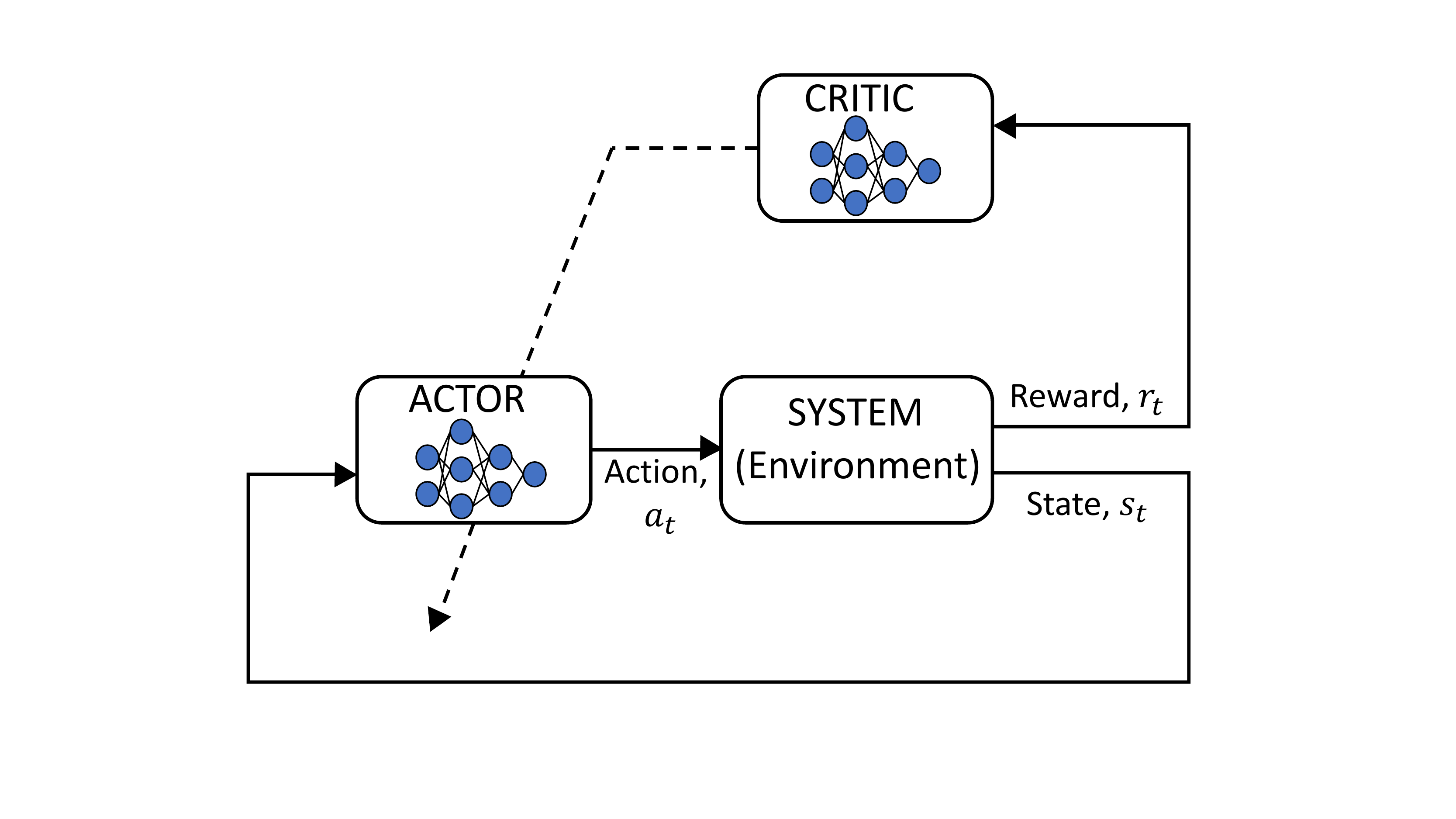}
\vspace{-4ex}
\caption{Actor-Critic Structure.}
\vspace{-4ex}
\label{fig:AC_framework}
\end{figure}

\subsection{Actor-Critic}


In RL, the action is taken by a \textit{policy} to maximize the total accumulated reward. By following a given policy and processing the rewards, one should estimate the expected return given states from a \textit{value function}. In the actor-critic approach, the \textit{actor} improves the policy based on the value function that is estimated by the \textit{critic} as depicted in Fig.~\ref{fig:AC_framework}. We specifically focus on the policy gradient-based actor-critic algorithm in this work, and, in particular, on the deep deterministic policy gradient (DDPG) \cite{lillicrap2015continuous}. This algorithm is an extension of deep Q-network (DQN) \cite{mnih2015human} to continuous actions, maintaining the importance of features such as: (i) random sampling from replay buffer where tuples are saved, (ii) the presence of target networks for stabilizing the learning process. The algorithm begins with a parameterized critic network, $Q(\bm{s}_{t},\bm{a}_{t} | \bm{\theta}^{Q})$, and actor network, $\pi(\bm{s}_{t} | \bm{\theta}^{\pi})$.


\subsubsection{Critic}

The role of the critic is to evaluate the current policy prescribed by the actor. Action is taken from the actor network with exploration noise, namely
\begin{align}
    \bm{a}_{t} = \pi(\bm{s}_{t} | \bm{\theta}^{\pi}) + \mathcal{N}_{t}
    \label{eqn:ddpg_action}
\end{align}
where $\pi(\bm{s}_{t})$ is a neural network, and $\mathcal{N}_{t}$ is exploration noise. After applying an action, we observe reward $\bm{r}_{t+1}$ and next state $\bm{s}_{t+1}$. The tuple $(\bm{s}_{t},\bm{a}_{t},r_{t+1},\bm{s}_{t+1})$ is stored in the replay buffer. We sample a random mini-batch of $N$ transitions from the buffer and for $i=1,\,\cdots,\,N$ we set
\begin{align}
    y_{i} = r_{i+1} + \gamma Q^{\prime}(\bm{s}_{i+1}, \pi^{\prime}(\bm{s}_{i+1} | \bm{\theta}^{\pi^{\prime}}) | \bm{\theta}^{{Q}^{\prime}})
\end{align}
where superscript $^{\prime}$ denotes the target network, whose parameters are slowly updated in order to track and filter the ones of the actual network thus reducing the chattering due to the learning process and enhancing its convergence. The critic network is updated to minimize the loss, $\mathcal{L}$:
\begin{align}
    \mathcal{L} & = \frac{1}{N}\sum_{i}\left(y_{i} - Q(\bm{s}_{i},a_{i} | \bm{\theta}^{Q})\right)^{2} \\
    \bm{\theta}^{Q}_{k+1} & = \bm{\theta}^{Q}_{k}+\eta_{Q} \nabla_{\bm{\theta}^{Q}} \mathcal{L}
\end{align}
where index-$k$ denotes the gradient descent algorithm iterates, and $\eta_{Q}$ denotes the learning rates of the critic network.
\subsubsection{Actor}
The parameters of the actor network are updated in order to maximize the long-term expected reward $\mathcal{J}(\theta^\pi)=V^\pi(\bm{s}_t)$ over episodes
\begin{align}\label{eq:update_actor}
    \bm{\theta}^{\pi}_{k+1} & = \bm{\theta}^{\pi}_{k} - \eta_{\pi} \nabla_{\bm{\theta}^{\pi}}\mathcal{J}
\end{align}
where index-$k$ denotes the gradient descent algorithm iterates, and $\eta_{\pi}$ denotes the learning rates of the actor network. Notice that according to the proof in \cite{silver2014deterministic}, the policy gradient in \eqref{eq:update_actor}   can be expressed as
\begin{align}
    \nabla_{\bm{\theta}^{\pi}}\mathcal{J}\approx \mathbb{E}_{\bm{s}_{t}\sim E}\left[\nabla_{\bm{a}}Q(\bm{s},\bm{a} | \bm{\theta}^{Q}) \rvert_{\bm{s}_{t},\,\pi(s_{t})} \nabla_{\bm{\theta}^{\pi}} \pi(\bm{s}|\bm{\theta}^{\pi})\rvert_{\bm{s}_{t}}  \right]
\end{align}
which is then approximated by samples as follows
\begin{align}
    \nabla_{\bm{\theta}^{\pi}}\mathcal{J} &\approx \frac{1}{N} \sum_{i} \nabla_{\bm{a}}Q(\bm{s},\bm{a} | \bm{\theta}^{Q}) \rvert_{\bm{s}_{i},\,\pi(s_{i})} \nabla_{\bm{\theta}^{\pi}} \pi(\bm{s}|\bm{\theta}^{\pi})\rvert_{\bm{s}_{i}}
    \end{align}

 Once the parameters of critic and actor network given samples are updated, then the target network is also updated as follows:
\begin{align}
    \bm{\theta}^{Q^{\prime}} &\leftarrow \tau \bm{\theta}^{Q} + (1 - \tau)\bm{\theta}^{Q^{\prime}} \nonumber \\
    \bm{\theta}^{\pi^{\prime}} &\leftarrow \tau \bm{\theta}^{\pi} + (1 - \tau)\bm{\theta}^{\pi^{\prime}}
    \label{eqn:target_update}
\end{align}
where $\tau$ is the level of ``soft-update''. Equation \eqref{eqn:target_update} improves the stability of the learning procedure. Note that convergence is no longer guaranteed, in general, when a value function approximator is used. Since the convergence of the critic network is not guaranteed, it is important to note that these target networks should update slowly to avoid divergence. Thus, one should choose a small value of $\tau$. This is a challenging point when the action space becomes continuous unlike tabular Q-learning.





\vspace{-1ex}
\section{Battery Charging problem}
\label{sec:BattModel}



In this section, we briefly discuss the battery models and control problem formulation used for the RL framework. We consider reduced order of electrochemical model that contains a large number of states, but achieves high-accuracy and represents physical details of battery dynamics. We also introduce the battery charging control problem formulation in this section.


\subsection{Electrochemical Model}
The Single Particle Model with Electrolyte and Thermal Dynamics (SPMeT) is derived from the Doyle-Fuller-Newman (DFN) electrochemical battery model. The DFN model employs a continuum of particles in both the anode and cathode of the cell. The SPMeT uses a simplified representation of solid phase diffusion that employs a single spherical particle in each electrode. The governing equations for SPMeT include linear and quasiliniar partial differential equations (PDEs) and a strongly nonlinear voltage output equation, given by:
\begin{align}
    \frac{\partial c_{s}^{\pm}}{\partial t}(r,t) &= \frac{1}{r^{2}} \frac{\partial}{\partial r} \left[ D_{s}^{\pm} r^{2} \frac{\partial c_{s}^{\pm}}{\partial r}(r,t) \right], \label{eqn:cs} \\
    \varepsilon_{e}^{j} \frac{\partial c_{e}^{j}}{\partial t}(x,t) &= \frac{\partial}{\partial x} \left[D_{e}^{\text{eff}}(c_{e}^{j}) \frac{\partial c_{e}^{j}}{\partial x}(x,t) + \frac{1 - t_{c}^{0}}{F} i_{e}^{j}(x,t) \right], \label{eqn:elec}
\end{align}
where $t \in \real_{+}$ represents time. The state variables are lithium concentration in the active particles of both electrode denoted by $c_{s}^{\pm}(r,t)$ and lithium concentration in the elctrolyte denoted by $c_{e}(x,t)$. $D_{s}^{\pm}$ and $D_{e}^{eff}(\cdot)$ are diffusion coefficients for solid phase and liquid phase dynamics. Note that superscript $j$ denotes anode, seperator and cathode, $j\in \{+,\textrm{sep},-\}$. Input current $I(t)$ is applied to the boundary conditions of governing PDEs. The terminal voltage output is governed by a combination of electric overpotential, electrode thermodyanmics, and Butler-Volmer kinetics, yielding:

\begin{align}\label{eqn:V}
\begin{split}
    V_{\text{T}}(t) =& \frac{RT_{cell}(t)}{\alpha F} \sinh^{-1}\left(\frac{-I(t)}{2a^+ A L^+ \bar{i}_0^+(t)}\right) \\&
    -\frac{RT_{cell}(t)}{\alpha F} \sinh^{-1}\left(\frac{I(t)}{2a^- A L^- \bar{i}_0^-(t)}\right) \\&
    + U^+\left(c_{ss}^+(t)\right)-U^-\left(c_{ss}^-(t)\right) \\&
    -\bigg{(}\frac{R_f^+}{a^+ A L^+}+\frac{R_f^-}{a^- A L^-}\bigg{)}I(t)\\&
    - \left(\frac{L^+ + 2L^{sep}+L^-}{2A\bar{\kappa}^{eff}}\right)I(t)\\&
    + k_{conc}(t)[ln(c_e(0^+,t))-ln(c_e(0^-,t))],
    \end{split}
\end{align}
where $c_{ss}$ is the solid phase surface concentration, namely $c_{ss}^{\pm}(x,t) = c_{s}^{\pm}(x,R_{s}^{\pm},t)$, $U^{\pm}$ is the open-circuit potential, and $c_{s,\max}^{\pm}$ is the maximum possible concentration in the solid phase. The nonlinear temperature dynamics are modeled with a simple heat transfer equation given by:
\begin{align}
    \frac{dT_{\text{cell}}}{dt}(t) = \frac{\dot{Q}(t)}{m c_{p;th}}-\frac{T_{\text{cell}}(t) - T_{\infty}}{m c_{p,th} R_{th}}
    \label{eqn:thermal}
\end{align}
where $T_{\text{cell}}$ represents cell temperature, $T_\infty$ is the ambient temperature, $m$ is the mass of the cell, $c_{p,th}$ is the thermal specific heat capacity of the cell, $R_{th}$ is the thermal resistance of the cell, and $\dot{Q}(t)$ is the heat added from the charging, which is governed by 
\begin{equation}\label{eqn:Qdot}
    \dot{Q}(t)=I(t)((U^{+}(SOC_p)-U^{-}(SOC_n))-V_{\text{T}}(t))
\end{equation}
with  the  convention  that  a  negative  current  is  charging  current, and $V(t)$ is the voltage determined by (\ref{eqn:V}). Both nonlinear open circuit potential functions in \eqref{eqn:Qdot} are functions of the bulk SOC in the anode and cathode, respectively. This heat generation term makes the temperature dynamics nonlinear. In this work, we focus on the SOC in anode expressed as a normalized volume sum along the radial axis:
\begin{align}
    SOC_{n} = \frac{3}{c_{s,max}^{-} (R_{s}^{-})^{3}}\int_{0}^{R_{s}^{-}} r^{2}c_{s}^{-}(r,t) dr.
    \label{eqn:soc}
\end{align}
For more details on the SPMeT equations, boundary conditions, and notations refer to \cite{Moura2017-SPMeObs,perez2017optimalJES}.
\vspace{-2ex}

\subsection{Minimum time charging problem}
The minimum time charging problem is formulated as:
\begin{align}
    \min_{I(t)} & \quad t_f \label{eqn:Obj_MinTime} \\ 
    \text{subject to} & \nonumber \\
    & \text{battery dynamics in \eqref{eqn:cs}-\eqref{eqn:soc} } \nonumber\\
    & V_T(t_0)=V_0,\, T_{\text{cell}}(t_0)=T_0\nonumber \\
    & SOC_{n}(t_f)=SOC_{\text{n,ref}},\, I(t) \in \left[I^{\text{min}},\,I^{\text{max}}\right] \nonumber \\
    & V_T(t) \leq V_T^{\text{max}},\, T_{\text{cell}}(t) \leq T_{\text{cell}}^{\text{max}} \nonumber
\end{align}
where $t_0=0$ and $t_f$ are the initial and final time of the charging procedure, $V_0$ and $T_0$ are the initial value for voltage and temperature respectively, $SOC_{\text{ref}}$ is the reference SOC at which the charging is considered complete. Moreover, $\left[I_{\text{min}},\,I_{\text{max}}\right]$ is the bound interval for the current while $V_T^{\text{max}}$ and $T_{\text{cell}}^{\text{max}}$, are the upper bounds for voltage and temperature. 
This is a free-time problem, whose objective is to solve the battery charging problem in minimum time given a battery model and operating constraints. Several publications use this formulation, including \cite{klein2011optimal, perez2017optimal, perez2017optimalJES}. This fast charging problem can be expanded in other forms by modifying the cost function in \eqref{eqn:Obj_MinTime} as maximize the charge throughput over specified time horizon. Work related to this formulation includes \cite{suthar2014optimal, suthar2014optimalJES}. On the other hand, authors in \cite{zou2018electrochemical, liu2017extended, aaron2020nn} consider SOC reference tracking problem where the cost function \eqref{eqn:Obj_MinTime} is defined as squared difference between the current SOC at time step $t$ and the reference SOC. These formulations fall within the class of state reference tracking problems.

\vspace{-1ex}
\section{Simulation Results}
\label{sec:Sim_results}

In this section, we conduct a case study on how the RL framework can be applied to the battery charging problem in simulation. Our goal is to obtain a charging control policy that charges the battery from 0.3 SOC to 0.8 SOC, while the states and outputs do not violate the constraints. We examine the performance of the actor-critic framework for the \textit{minimum time charging} problem using the electrochemical model in Section \ref{sec:BattModel}. When an electrochemical model is considered, ADP methods become a sensible choice due to the large number of states. We first assume that all the  states are available to the agent. Then, we drop this assumption and consider the more realistic scenario in which only temperature and SOC can be measured/computed. Furthermore, we are interested in seeing how actor-critic adapts its learning behavior when the environment changes. This is especially important in battery applications, where the optimal charging trajectory will vary as the battery ages.





In this case study, we consider the minimum charging problem for an electrochemical model, whose chemistry is based on graphite anode/LiNiMnCoO2 (NMC) cathode cell. The PDEs in \eqref{eqn:cs}-\eqref{eqn:elec} are spatially discretized by finite difference. Then, state-space representation is formed with these discretized states and thermal state \eqref{eqn:thermal} , which results in a relatively large-scale dynamical systems, 61 states. The actor-critic networks are based on neural network architectures \cite{lillicrap2015continuous} with different numbers of neurons. Specifically, the actor network uses two hidden layers with 20 - 20 neurons. The critic network uses two hidden layers with  100 - 75 neurons. Hyper parameters are detailed in Table~\ref{tbl:ActorCriticParam}.

\begin{table}[h!]
\centering
\scalebox{1.3}{
\begin{tabular}{ |c|c|c| }
 \hline
 Variable & Description &Value \\
 \hline
 $\gamma$ & Discount factor & $0.99$ \\
 $\eta_{\pi}$, $\eta_{Q}$ & Learning rate of actor, critic & $10^{-4}$, $10^{-3}$ \\
 $\tau$ & Soft update of target networks &$10^{-3}$ \\
 \hline
\end{tabular}
}
\caption{Actor-critic hyper parameters.}
\vspace{-3ex}
\label{tbl:ActorCriticParam}
\end{table}
The reward function is designed with the aim of both achieving fast charging and guaranteeing safety, according to the optimization problem in \eqref{eqn:Obj_MinTime}
\begin{align}
    r_{t+1} = r_{\text{fast}} + r_{\text{safety}}(\bm{s}_t,\bm{a}_t)
\end{align}
where $r_{\text{fast}}=-0.1$ is a negative penalty for each time step which passes before the reference SOC is achieved. In addition, a negative penalty is also introduce at each time step in which the voltage and temperature constraints are violated 
\begin{align}
    r_{\text{safety}}(\bm{s}_t,\bm{a}_t)=r_{\text{volt}}(\bm{s}_t,\bm{a}_t)+r_{\text{temp}}(\bm{s}_t,\bm{a}_t)
\end{align}
This is done in particular by means of  linear penalty functions \cite{smith1995penalty}:
\begin{align}
    r_{\text{volt}}(\bm{s}_t,\bm{a}_t) & = \begin{cases}
				-100(V_{\text{T}}(t) - V_{\text{T}}^{\text{max}}), & \text{if}\ V_{\text{T}}(t) \geq V_{\text{T}}^{\text{max}} \label{eq:r_volt} \\
				0, & \text{otherwise}
				\end{cases}  \\
	r_{\text{temp}}(\bm{s}_t,\bm{a}_t) & = \begin{cases}
	            -5(T_{\text{cell}}(t) - T_{\text{cell}}^{\text{max}}), & \text{if}\ T_{\text{cell}}(t) \geq T_{\text{cell}}^{\text{max}} \label{eq:r_temp} \\
	            0, & \text{otherwise}
	            \end{cases} \\ \nonumber
\end{align}
where constraints are set to $V_{\text{T}}^{\text{max}}=4.2V$, $T_{\text{cell}}^{\text{max}} = 47^{\circ}C$ in this case study. The current is limited within the range $[0, 1.8C]$, where $C$ is the C-rate related to the considered cell. The current is applied by scaling and translating the output of the actor network which, in the considered case, is already limited in the range $[-1,\,1]$, due to the fact that its last layer is an hyperbolic tangent operator, i.e., $-1 \leq \tanh(\cdot) \leq 1$.
\vspace{-2ex}

\subsection{Learning Constrained Charging Controls}
\label{subsubsec:Constrained}

\begin{figure*}[]
      \centering  
      \includegraphics[trim = 0mm 0mm 0mm 0mm, clip, width=0.9\textwidth]{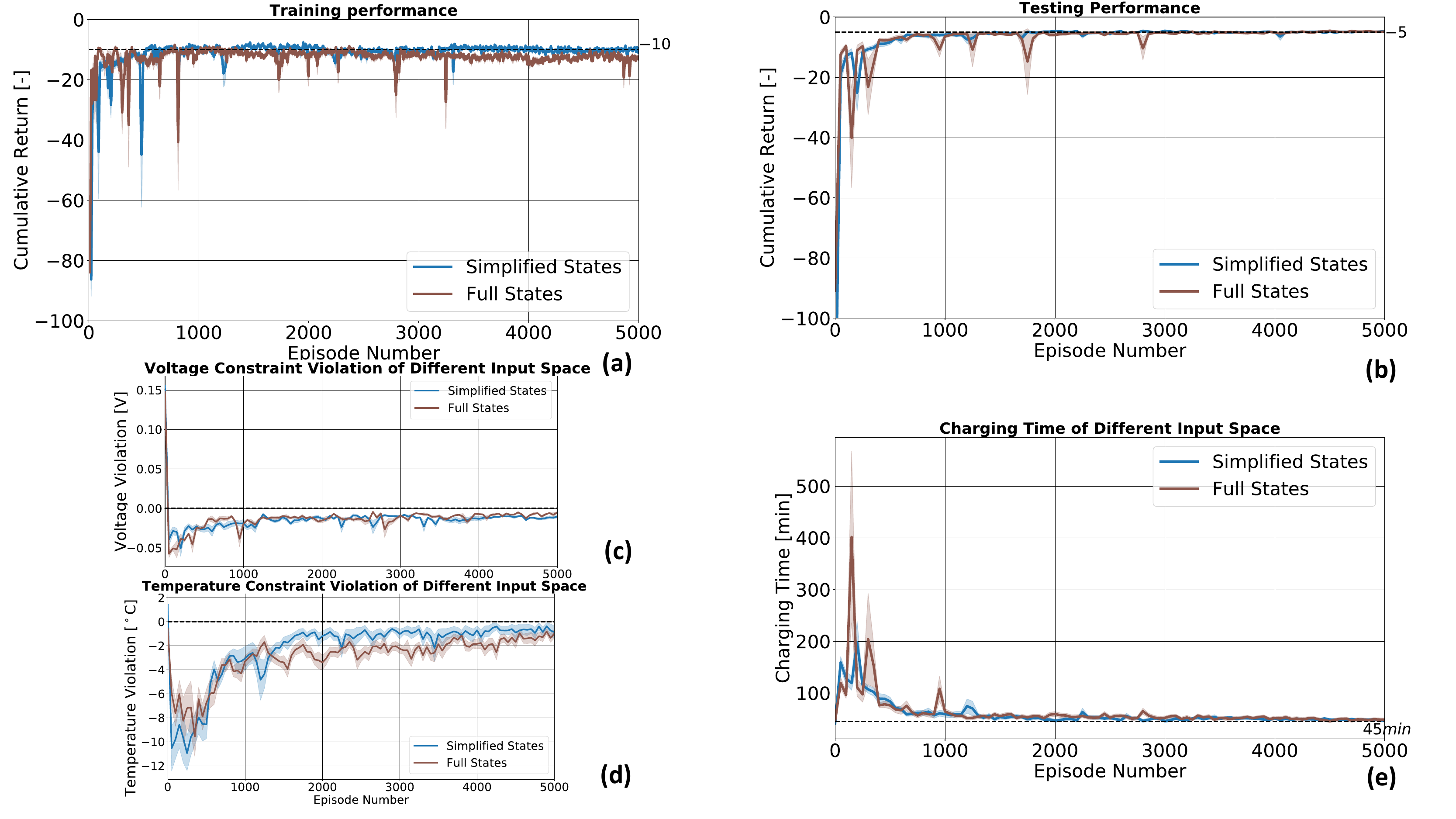}
      \caption{Actor-critic \textit{constrained} charging results for the SPMeT model with 95 \% confidence interval. (a) training performance, (b) testing performance, (c,d) constraint violation, (e) charging time}
      \label{fig:CaseStudy2a}
\vspace{-4ex}
\end{figure*}

\begin{figure}[t!]
\centering
\includegraphics[trim = 15mm 15mm 20mm 10mm, clip, width=0.9\linewidth]{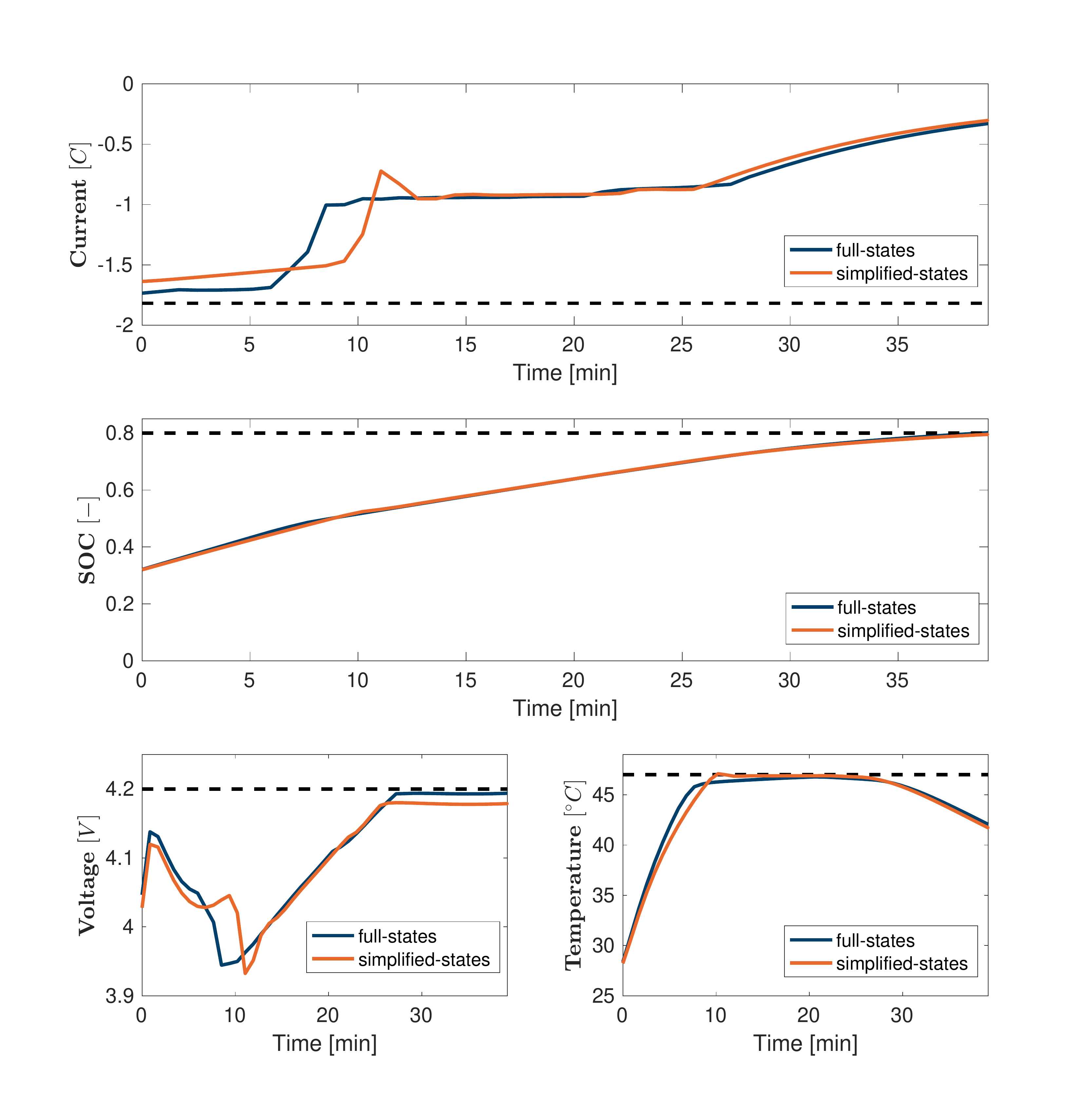}
\vspace{-2ex}
\caption{Validation of actor-critic algorithm after training with $V_{\text{T}}(0)=3.6V, T_{\text{cell}}(0)=27^{\circ}C$: (a) simplified-states achieves -5.38 cumulative rewards; (b) full-states achieves -4.69 cumulative rewards.}
\vspace{-2ex}
\label{fig:CaseStudy2a_valid}
\end{figure}

The objective of this study is to: (i) validate the actor-critic performance on the minimum time charging problem, and (ii) compare the performance with full and reduced state feedback for the actor-critic networks. The performance is measured by the cumulative reward for each episode. In training, the action is determined by following \eqref{eqn:ddpg_action} with the presence of exploration noise. In testing, we test the policy without exploration noise so that we can see the performance of the trained actor-critic network.

Figures~\ref{fig:CaseStudy2a}a-b show the training/testing results of the actor-critic approach. The performance of controller during training converges to around -10 cumulative reward while the performance of controller during testing converges to around -5 cumulative reward. The difference comes from the presence of exploration noise. We can clearly observe that exploration is not needed after 1000 episodes as the action network, $\pi$, falls into its local optimal. Furthermore, we design two state-feedback controllers\footnote{Note the state vectors are inputs for both the actor and critic networks}. One utilizes the full state vector (61 states) for feedback control. The other uses a reduced or ``simplified'' state vector for feedback control, with only SOC and temperature (2 states). The purpose of reducing the state vector size is motivated from an intuition that the objective function only involves anode bulk SOC and state constraints. The training/testing results in Fig.~\ref{fig:CaseStudy2a}a-b show that both simplified-states and full-states achieve the goal.

Figures~\ref{fig:CaseStudy2a}c-d show how much constraints are violated during testing. The constraint violation scores are calculated according to $\max\left\{ V_{\text{T}}(t) - V_{\text{max}} \, \forall t \in \left[ 0, t_f \right] \right\}$, $\max \left\{ T_{\text{cell}}(t) - T_{\text{cell}}^{\text{max}} \, \forall t \in \left[0, t_f \right] \right\}$ for each episode. The constraint violation scores approach zero as the episodes increase, which implies that the controller learns the constraints. Positive values imply the constraints are violated. The constraints are violated in the beginning because parameters of actor-critic are randomly initialized. However, they approach the boundary during learning, since the optimal solution is along the constraint boundary. Figure~\ref{fig:CaseStudy2a}e shows that the charging time decreases to about 45 minutes. Figure~\ref{fig:CaseStudy2a_valid} visualizes the action, states, and constraints for the simplified-states and full-states actor-critic networks. We find that both controllers achieve similar performance for minimum time charging, around 40 minutes for the given initial conditions. The derived feedback control policy exhibits the constant current (CC), constant temperature (CT), and constant voltage (CV) shape, which can be qualitatively similar to the model-based control results in \cite{klein2011optimal,torchio2015real, perez2017optimal, aaron2020nn, zou2018electrochemical}. The only difference is that we don't require any knowledge of model dynamics.


\vspace{-2ex}
\subsection{Learning Adaptive Constrained Charging Controls}
\label{subsubsec:adaptive}

\begin{figure*}[]
      \centering  
      \includegraphics[trim = 0mm 0mm 0mm 0mm, clip, width=0.9\textwidth]{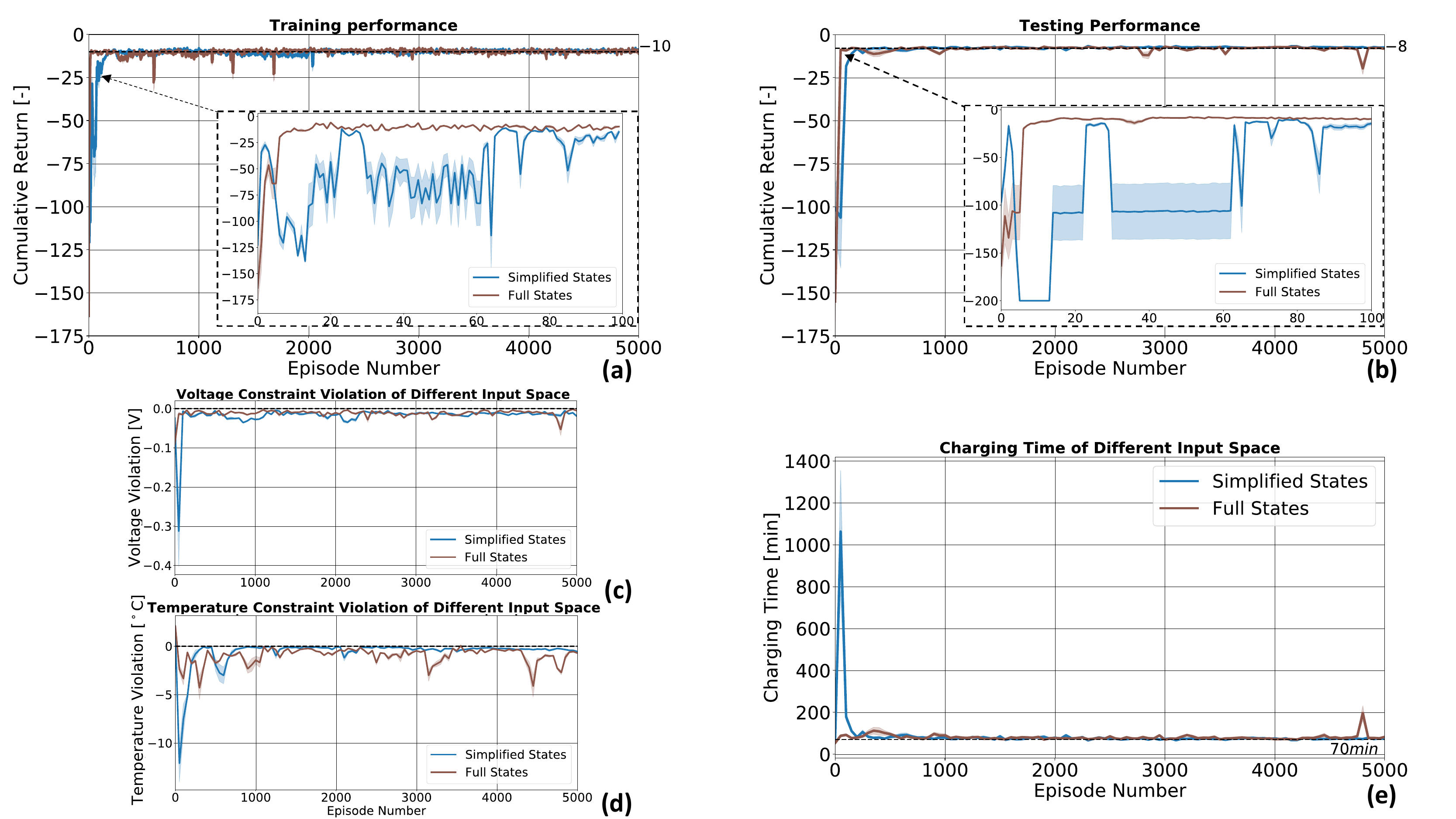}
      \caption{Actor-critic \textit{adaptive} charging results for the SPMeT model with 95 \% confidence interval. (a) training performance, (b) testing performance, (c,d) constraint violation, (e) charging time}
      \vspace{-4ex}
      \label{fig:CaseStudy2b}
\end{figure*}

\begin{figure}[t!]
\centering
\includegraphics[trim = 15mm 15mm 20mm 10mm, clip, width=0.9\linewidth]{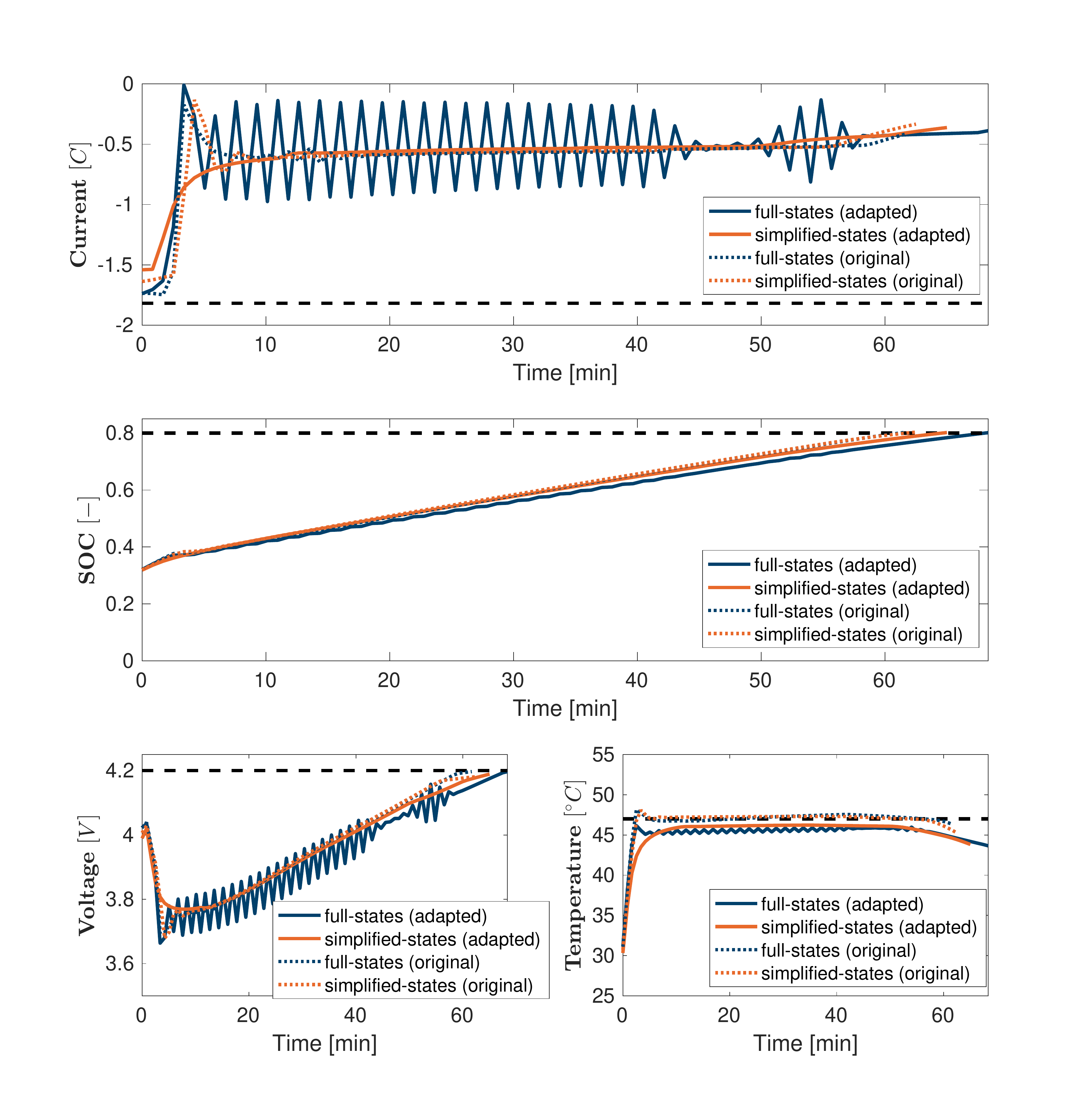}
\vspace{-2ex}
\caption{The validation of \textit{adaptive} actor-critic algorithm after training at $V_{\text{T}}(0)=3.6V, T_{\text{cell}}(0)=27^{\circ}C$: (a) \textit{adapted} simplified-states achieves -7.79 cumulative rewards while original simplified-states achieves -81.78 due to state violation; (b) \textit{adapted} full-states achieves -8.19 cumulative rewards while original full-states achieves -82.16 due to state violation.}
\vspace{-4ex}
\label{fig:CaseStudy2b_valid}
\end{figure}

In this section, we are interested in the adaptability of the actor-critic approach, which is crucial to the battery charging problem as the cell ages. To represent aging, we perturb the electrochemical parameters, namely, film resistance, $R_f^{\pm}$, heat generation, $\dot{Q}(t)$. Perturbation of those parameters represents the battery degradation as they directly affect to battery voltage \eqref{eqn:V} and thermal state \eqref{eqn:thermal}, which can be monitored by experimental measurement. We expect that the previous actor-critic network could violate the state constraints immediately.

Figures~\ref{fig:CaseStudy2b}a-b display the adaptation results of the actor-critic approach in training/testing. We start from the previous actor-critic configuration in order to observe its adaptability. We observe that both full-sates and simplified actor-critic network are capable of adapting its policy to achieve the goal. We take zoom-in the first 100 episodes to see how the adaptation is processed for the episodes. We observe that the full-states actor-critic network adapts much faster than the simplified states. This is related to the large number of parameters in the full-states network which can lead to greater flexibility in adapting to the new environment.

Figures~\ref{fig:CaseStudy2b}c-d describe the constraint violation scores. Due to change of environment, we observe that the controllers are prone to violate the constraints. Figure~\ref{fig:CaseStudy2b}e shows that the battery charging time increases compared to previous case study because of state violations. The controller reduces the charging level thus increasing the required charging time in order  to reach the reference SOC (0.8) from 45 to 70 minutes.

Figure~\ref{fig:CaseStudy2b_valid} shows the performance of adaptive controllers at the end of training. The constraints are equivalent to the previous validation, but the system dynamics has changed. So, the original actor-critic network, which learns from a fresh battery, immediately violates the constraints since the environment has changed (aged). However, we are able to construct adaptive controllers for both full-states and simplified-states that achieve the goal without safety violations from previous actor-critic networks. The fluctuating current for the full-states actor-critic network could be mitigated by regularizing the actor-critic parameters during learning.\vspace{-2ex}


\vspace{-1ex}
\section{Conclusion}
\label{sec:Conclusion}
\vspace{-0ex}
In this paper, we have examined a reinforcement learning approach for the battery fast-charging problem in the presence of safety constraints. In particular, we have shown how RL can overcome many of the limitations of the model-based methods. Among the RL paradigms, the actor-critic paradigm, and specifically the DDPG algorithm, has been adopted due to its ability to deal with continuous state and action spaces. To address the state constraints, the reward function has been designed such that the agent learns constraint violation. The control strategy has been tested in simulation on an electrochemical battery model and the presented results are consistent with model-based approaches. In addition, the performance of the actor-critic strategy has been evaluated both in the case of full and partial state feedback. Finally, the adaptability of the control algorithm to battery ageing has been considered. Future work involves adding different types of safety constraints, related to electrochemical phenomena occurring inside the battery, and experimental validation.

\vspace{-1ex}
\section{Acknowledgements}
This research is funded by LG Chem Battery Innovative Contest. The authors thank the LG Chem researchers for their support and discussion in the work.

\ifCLASSOPTIONcaptionsoff
  \newpage
\fi



%


\bibliographystyle{unsrt}
\bibliography{Sample}



%





\end{document}